\documentclass[twocolumn,showpacs,superscriptaddress,
preprintnumbers,floatfix]{revtex4}
\usepackage{graphicx,epsf,amssymb,amsmath,latexsym}
\newcommand{\be}{\begin{eqnarray}}
\newcommand{\ee}{\end{eqnarray}}

\renewcommand{\d}{\mbox{${\rm d}$}}

\begin{document}
\title{Improved WKB analysis of Slow-Roll Inflation}
\author{Roberto Casadio}
\email{Roberto.Casadio@bo.infn.it}
\affiliation{Dipartimento di Fisica, Universit\`a di Bologna and I.N.F.N.,
Sezione di Bologna, via~Irnerio~46, 40126 Bologna, Italy.}
\author{Fabio Finelli}
\email{finelli@bo.iasf.cnr.it}
\affiliation{IASF/INAF, Istituto di Astrofisica Spaziale e Fisica Cosmica,
Sezione di Bologna,
Istituto Nazionale di Astrofisica,
via~Gobetti~101, 40129 Bologna, Italy.}
\author{Mattia Luzzi}
\email{Mattia.Luzzi@bo.infn.it}
\affiliation{Dipartimento di Fisica, Universit\`a di Bologna and I.N.F.N.,
Sezione di Bologna, via~Irnerio~46, 40126 Bologna, Italy.}
\author{Giovanni Venturi}
\email{armitage@bo.infn.it}
\affiliation{Dipartimento di Fisica, Universit\`a di Bologna and I.N.F.N.,
Sezione di Bologna, via~Irnerio~46, 40126 Bologna, Italy.}
\begin{abstract}
We extend the WKB method for the computation of cosmological perturbations
during inflation beyond leading order and provide the power spectra of scalar
and tensor perturbations to second order in the slow-roll parameters.
Our method does not require that the slow-roll parameters be constant.
Although leading and next-to-leading results in the slow-roll parameters
depend on the approximation technique used in the computation, we find
that the inflationary theoretical predictions obtained may
reach the accuracy required by planned observations.
In two technical appendices, we compare our techniques and results with
previous findings.
\end{abstract}
\maketitle
\section{Introduction}
\label{intro}
One of the main topics in theoretical cosmology is the computation of
the spectra of perturbations during inflation (for an introduction
and review see, e.g.~Refs.~\cite{infla}), which are then to be compared
with the data for the Cosmic Microwave Background Radiation (CMBR)
obtained in present and future experiments~\cite{wmap,planck}.
It is in general impossible to obtain such spectra analytically,
and therefore approximate methods have been proposed, amongst which
stands out the slow-roll approximation of Ref.~\cite{SL}.
This amounted to introducing a suitable hierarchy of parameters,
whose time-dependence was frozen to first order, and corresponding
spectra and spectral indices were then determined uniquely.
\par
The task of extending standard slow-roll results beyond first
order is however highly non-trivial.
As was noted in Ref.~\cite{wang_mukh}, one expects that
the time-dependence of the parameters that describe the
inflationary Universe could then appear in the order of the
Bessel functions which describe the perturbations.
Since this is, in general, inconsistent, other ways have been
devised in order to by-pass this issue, such as the
Green's function method of Refs.~\cite{SGong,LLMS}, and
the ``uniform'' approximation developed in
Refs.~\cite{H_MP,HHHJM,HHHJ}.
\par
Our approach to this problem is to make use of an improvement
of the WKB approximation (first applied to cosmological
perturbations in Ref.~\cite{MS_WKB})
introduced in Ref.~\cite{WKB1}.
We followed the method of Ref.~\cite{langer} in order to
illustrate an improved (uniform) leading order approximation
and then introduced two expansions:
the ``adiabatic'' expansion of Ref.~\cite{langer} and a
perturbative Green's function expansion.
In the present paper, we shall use our methods
(in particular the adiabatic expansion of Ref.~\cite{WKB1})
in conjunction with the slow-roll approximation~\cite{SL,MS}
in order to obtain higher order terms in the slow-roll parameters.
Our method will not require that the slow-roll parameters be
constant and extends the results illustrated in Ref.~\cite{WKB_PRL}.
Let us emphasize here the distinction between the slow-roll and WKB
formalisms.
The latter (also referred to as ``semiclassical'' or adiabatic expansion)
is used to obtain approximate solutions to the general cosmological
perturbation equations, whereas the former corresponds to an
approximate solution for the homogeneous mode in which the inflaton
kinetic term is neglected and one has a hierarchy of horizon flow
functions (see Section~\ref{basics}).
Thus, they correspond to diverse, and independent, approximation
schemes, even if, as is obvious, the validity (or not) of the slow-roll
approximation will influence the form of the approximate WKB solution.
\par
In Section~\ref{basics}, we briefly recall the main definitions
and standard slow-roll results.
Our results to leading WKB and second slow-roll order,
and next-to-leading WKB and first slow-roll order from
Ref.~\cite{WKB_PRL} are respectively reviewed in detail in
Sections~\ref{1+2} and \ref{2+1}.
New results for the next-to-leading WKB and second slow-roll
order are given in Section~\ref{2+2}.
We finally comment on our work in Section~\ref{conc} and compare
with other results in the literature in Appendices~\ref{app} and
\ref{app_other_res}.
\section{Basics}
\label{basics}
Let us begin by recalling the Robertson-Walker metric in
conformal time $\eta$,
\be
\d s^2=a^2(\eta)\,\left[-\d\eta^2+\d\vec x\cdot \d\vec x\right]
\ ,
\ee
where $a$ is the scale factor of the Universe.
Scalar (density) and tensor (gravitational wave) fluctuations
are given on this background, respectively by
$\mu=\mu_{\rm S}\equiv a\,Q$ ($Q$ is the Mukhanov
variable~\cite{mukh}) and $\mu=\mu_{\rm T}\equiv a\,h$
($h$ is the amplitude of the two polarizations
of gravitational waves~\cite{gris,staro}), where
the functions $\mu$ must satisfy the one-dimensional
Schr\"odinger-like equation
\be
\left[\frac{{\d}^2}{{\d}\eta^2}+\Omega^2(k,\eta)\right]
\,\mu=0
\ ,
\label{osci}
\ee
together with the initial condition
\be
\lim_{\frac{k}{a\,H}\rightarrow +\infty} \mu(k,\eta)
\simeq\frac{{\rm e}^{-i\,k\,\eta}}{\sqrt{2\,k}}
\ .
\label{init_cond_on_mu}
\ee
In the above $k$ is the wave-number,
$H\equiv a'/a^2$ is the Hubble parameter,
and
\be
\Omega^2(k,\eta)\equiv k^2-\frac{z''}{z}
\ ,
\label{freq}
\ee
where $z=z_{\rm S}\equiv a^2\,\phi'/H$ for scalar
and $z=z_{\rm T}\equiv a$ for tensor perturbations
($\phi$ is the homogenous inflaton and primes denote
derivatives with respect to $\eta$).
The dimensionless power spectra (PS henceforth) of scalar
and tensor fluctuations are then given by
\begin{subequations}
\be
\mathcal{P}_{\zeta}\equiv
\displaystyle\frac{k^{3}}{2\,\pi^{2}}\,
\left|\frac{\mu_{\rm S}}{z_{\rm S}}\right|^{2}
\ ,
\ \ \ \
\mathcal{P}_{h}\equiv
\displaystyle\frac{4\,k^{3}}{\pi^{2}}\,
\left|\frac{\mu_{\rm T}}{z_{\rm T}}\right|^{2}
\label{spectra_def}
\ee
and the spectral indices and runnings
\be
n_{\rm S}-1\equiv
\displaystyle\frac{\d\ln \mathcal{P}_{\zeta}}
{\d\ln k}
\ ,
\quad
n_{\rm T}\equiv
\displaystyle\frac{\d\ln \mathcal{P}_{h}}
{\d\ln k}
\label{n_def}
\ee
\be
\alpha_{\rm S}\equiv
\frac{\d^{2}\ln\mathcal{P}_{\zeta}}
{(\d\ln k)^{2}}
\ ,
\quad
\alpha_{\rm T}\equiv
\frac{\d^{2}\ln \mathcal{P}_{h}}
{(\d\ln k)^{2}}
\ .
\label{alpha_def}
\ee
We also define the tensor-to-scalar ratio
\be
R\equiv\frac{\mathcal{P}_{h}}{\mathcal{P}_{\zeta}}
\ .
\label{R_def}
\ee
\end{subequations}
\par
The evolution of the Universe is usually described by means
of a set of flow equations~\cite{MS_WKB,terrero}.
The zero horizon flow function is defined by
\be
\epsilon_0\equiv\frac{H(N_{i})}{H(N)}
\ ,
\label{eps_0}
\ee
where $N$ is the number of e-folds, $N\equiv\ln (a/a_i)$
[where $a_i=a(\eta_i)$] after the arbitrary initial time
$\eta_i$.
The hierarchy of horizon flow functions (HFF henceforth)
is then defined according to
\be
\epsilon_{i+1}\equiv
\frac{\d\ln |\epsilon_i|}{\d N}
\ ,
\quad\quad
i\geq 0
\label{hor_flo_fun}
\ee
and inflation takes place for $\epsilon_1 < 1$ (see
Appendix~\ref{app} for a comparison with different
conventions).
We finally recall that the frequencies $\Omega$
can now be expressed in terms of $\epsilon_1$,
$\epsilon_2$, and $\epsilon_3$ only, as
\be
&&
\frac{k^2-\Omega^2_{\rm T}}{a^{2}\,H^{2}}=2-\epsilon_1
\nonumber
\\
\label{pot}
\\
&&
\frac{k^2-\Omega^2_{\rm S}}{a^{2}\,H^{2}}
=
2-\epsilon_1+\frac{3}{2}\,\epsilon_2
-\frac{1}{2}\,\epsilon_{1}\,\epsilon_2
+\frac{1}{4}\,\epsilon_2^2
+\frac{1}{2}\,\epsilon_2\,\epsilon_3
\ .
\nonumber
\ee
In general, $\Omega^2(k,\eta)$ vanishes for a certain
time $\eta=\eta_0(k)$ (the classical turning point).
\subsection{WKB Formalism}
\label{form}
Let us briefly summarize the essential formulae from our
Ref.~\cite{WKB1}.
First of all, we apply the Langer transformation
\be
x\equiv\ln\left(\frac{k}{H\,a}\right)
\ ,
\ \ \
\chi\equiv(1-\epsilon_1)^{1/2}\,{\rm e}^{-x/2}\,\mu
\label{transf}
\ee
so that Eq.~(\ref{osci}) becomes
\be
\left[\frac{\d^2}{\d x^2}+\omega^2(x)\right]\,\chi(x)=0
\label{new_eq}
\ ,
\ee
where for scalar and tensor perturbations one now has
\begin{widetext}
\be
\omega_{\rm S}^2(x)&=&
\frac{{\rm e}^{2\,x}}{(1-\epsilon_1)^2}
-\frac{1}{4}\,\left(\frac{3-\epsilon_1}{1-\epsilon_1}\right)^2
-\frac{(3-2\,\epsilon_1)\,\epsilon_2}{2\,(1-\epsilon_1)^2}
-\frac{(1-2\,\epsilon_1)\,\epsilon_2\,\epsilon_3}{2\,(1-\epsilon_1)^3}
-\frac{(1-4\,\epsilon_1)\,\epsilon_2^2}{4\,(1-\epsilon_1)^4}
\nonumber
\\
\label{omega2_ST}
\\
\omega_{\rm T}^2(x)&=&
\frac{{\rm e}^{2\,x}}{(1-\epsilon_1)^2}
-\frac{1}{4}\,\left(\frac{3-\epsilon_1}{1-\epsilon_1}\right)^2
+\frac{\epsilon_1\,\epsilon_2}{2\,(1-\epsilon_1)^2}
+\frac{\epsilon_1\,\epsilon_2\,\epsilon_3}{2\,(1-\epsilon_1)^3}
+\frac{(2+\epsilon_1)\,\epsilon_1\,\epsilon_2^2}{4\,(1-\epsilon_1)^4}
\ ,
\nonumber
\ee
\end{widetext}
and the corresponding turning points are mapped into
$x=x_0(\eta_0,k)$.
We then denote quantities evaluated in the region to
the right (left) of the turning point $x_0$ with the
subscript I (II).
For example,
\be
\omega_{\rm I}(x)=\sqrt{\omega^{2}(x)}
\ ,
\ \ \
\omega_{\rm II}(x)=\sqrt{-\omega^{2}(x)}
\ee
and
\be
\xi_{\rm I}(x)=\int_{x_0}^{x}\omega_{\rm I}(y)\,\d y
\ ,
\ \ \
\xi_{\rm II}(x)=\int_{x}^{x_0}\omega_{\rm II}(y)\,\d y
\ .
\label{xiI_II}
\ee
\par
The next-to-leading WKB solutions of Eq.~(\ref{new_eq}),
in the adiabatic expansion and for one {\em linear turning point},
is then given by a combination of Bessel functions and their
derivatives as~\cite{WKB1,langer}
\be
U_{\rm I}^{(\pm)}(x)
&\!\!\!\!=\!\!\!\!&
\left[1+\phi_{{\rm I} (1)}(x)\right]
\,u_{{\rm I}}^{(\pm)}(x)
+\gamma_{{\rm I} (1)}(x)\,u_{\rm I}^{(\pm)'}(x)
\nonumber
\\
\label{sol_NTL}
\\
U_{\rm II}^{(\pm)}(x)
&\!\!\!\!=\!\!\!\!&
\left[1+\phi_{{\rm II} (1)}(x)\right]
\,u_{{\rm II}}^{(\pm)}(x)
+\gamma_{{\rm II} (1)}(x)\,u_{\rm II}^{(\pm)'}(x)
\nonumber
\ ,
\ee
where
\be
&&
u_{{\rm I}}^{(\pm)}(x)=\sqrt{\frac{\xi_{\rm I}(x)}
{\omega_{\rm I}(x)}}\,
J_{\pm \frac{1}{3}}\left[\xi_{\rm I}(x)\right]
\nonumber
\\
\label{particular_solI}
\\
&&
u_{{\rm II}}^{(\pm)}(x)=\sqrt{\frac{\xi_{\rm II}(x)}
{\omega_{\rm II}(x)}}\,
I_{\pm \frac{1}{3}}\left[\xi_{\rm II}(x)\right]
\ .
\nonumber
\ee
The above expressions~(\ref{particular_solI}) satisfy
an equation of motion
\be
\left[\frac{\d^2}{\d x^2}+\omega^2(x)-\sigma(x)\right]\,\chi(x)=0
\label{eq_with_sigma}
\ ,
\ee
where the residual term is defined by
\be
\sigma(x)&\equiv&
\frac{3}{4}\,\frac{(\omega')^2}{\omega^2}
-\frac{\omega''}{2\,\omega}
-\frac{5}{36}\,\frac{\omega^{2}}{\xi^{2}}
\ .
\label{omega_xi_sigma}
\ee
As we showed in Ref.~\cite{WKB1}, such a quantity is expected
to be small for the cases of interest, that is, in the
{\em sub-horizon limit}
\be
\frac{k}{a\,H}={\rm e}^x\to +\infty
\ ,
\ee
in the {\em super-horizon limit}
\be
\frac{k}{a\,H}={\rm e}^x\to 0
\ ,
\ee
and also at the turning points (zeros) $x=x_0$ of the frequencies.
In Eq.~(\ref{omega_xi_sigma}), we omitted the subscripts I and II
for brevity and primes hereafter will denote derivatives with
respect to $x$.
\par
The next-to-leading WKB PS in the adiabatic
expansion are given by
\begin{subequations}
\be
&&
{\cal P}_\zeta=
\frac{H^2}{\pi\,\epsilon_1\,m_{\rm Pl}^2}
\left(\frac{k}{a\,H}\right)^3
\frac{{\rm e}^{2\,\xi_{\rm II,S}}\,\left(1+g_{(1){\rm S}}^{\rm AD}\right)}
{\left(1-\epsilon_1\right)\,\omega_{\rm II,S}}
\nonumber
\\
\label{spectra_correct}
\\
&&
{\cal P}_h=
\frac{16\,H^2}{\pi\,m_{\rm Pl}^2}
\,\left(\frac{k}{a\,H}\right)^3\,
\frac{{\rm e}^{2\,\xi_{\rm II,T}}\,\left(1+g_{(1){\rm T}}^{\rm AD}\right)}
{\left(1-\epsilon_1\right)\,\omega_{\rm II,T}}
\ .
\nonumber
\ee
The spectral indices are also given by
\be
&&
n_{\rm S}-1=
3+2\,
\frac{\d\,\xi_{\rm II,S}}{\d\,\ln k}
+
\frac{\d\,g_{(1){\rm S}}^{\rm AD}}{\d\,\ln k}
\nonumber
\\
\label{index_correct}
\\
&&
n_{\rm T}=
3+2\,
\frac{\d\,\xi_{\rm II,T}}{\d\,\ln k}
+
\frac{\d\,g_{(1){\rm T}}^{\rm AD}}{\d\,\ln k}
\ ,
\nonumber
\ee
and their runnings by
\be
&&
\alpha_{\rm S}=
2\,
\frac{\d^2\,\xi_{\rm II,S}}{\left(\d\,\ln k\right)^2}
+
\frac{\d^2\,g_{(1){\rm S}}^{\rm AD}}{\left(\d\,\ln k\right)^2}
\nonumber
\\
\label{runnings_correct}
\\
&&
\alpha_{\rm T}=
2\,
\frac{\d^2\,\xi_{\rm II,T}}{\left(\d\,\ln k\right)^2}
+
\frac{\d^2\,g_{(1){\rm T}}^{\rm AD}}{\left(\d\,\ln k\right)^2}
\ .
\nonumber
\ee
Finally, the tensor-to-scalar ratio takes the form
\be
R=16\,\epsilon_1\,
\frac{{\rm e}^{2\,\xi_{\rm II,T}}\,\left(1+g_{(1){\rm T}}^{\rm AD}\right)\,
\omega_{\rm II,S}}
{{\rm e}^{2\,\xi_{\rm II,S}}\,\left(1+g_{(1){\rm S}}^{\rm AD}\right)\,
\omega_{\rm II,T}}
\ ,
\label{R_correct}
\ee
\end{subequations}
where all quantities are evaluated in the super-horizon limit
($x\ll x_0$) and we used the results
\begin{subequations}
\be
g_{(1){\rm S,T}}^{\rm AD}(x)&\!\!\!=\!\!\!&
2\left\{
\phi_{{\rm II}(1)}(x)
-\gamma_{{\rm II}(1)}(x)
\left[\omega_{\rm II}(x)
+\frac{\omega_{\rm II}'(x)}{2\,\omega_{\rm II}(x)}\right]
\right.
\nonumber
\\
&&
\left.
\phantom{2\,[}
+\frac{1}{2}\,\gamma_{{\rm I}(1)}(x_i)
-\phi_{{\rm I}(1)}(x_i)
\right\}_{\rm S,T}
\label{g_1AD}
\ee
\be
\phi_{(1)}(x)&\!\!=\!\!&-\frac{1}{2}\,\int^{x}
\left[\gamma''_{(1)}(y)+\sigma(y)\,\gamma_{(1)}(y)\right]\,
\d y
\nonumber
\\
\label{phi_1}
\\
\gamma_{(1)}(x)&\!\!=\!\!&\frac{1}{2\,\omega(x)}\,\int^x
\frac{\sigma(y)}{\omega(y)}\,
\d y
\ ,
\nonumber
\ee
\end{subequations}
where it is understood that the integration must be performed
from $x_0$ to $x$ in region~I and from $x$ to $x_0$ in
region~II, and the indices S and T refer to the use of the
corresponding frequencies.
Let us note that if we take the above formulae with
$g_{(1){\rm S,T}}^{\rm AD}(x)\equiv0$, we obtain the leading
WKB results given in Ref.~\cite{MS_WKB} for the PS,
spectral indices and $\alpha$-runnings.
\subsection{Slow-Roll Results}
Let us briefly recall the known results obtained
from the slow-roll approximation~\cite{SL,MS_WKB,MS}.
To first order in the slow-roll parameters~(\ref{hor_flo_fun}),
we have the PS
\begin{subequations}
\begin{widetext}
\be
&&
\mathcal{P}_{\zeta}=\frac{H^2}{\pi\,\epsilon_1\,m_{\rm Pl}^2}
\,\left[1-2\,(C+1)\,\epsilon_1-C\,\epsilon_2
-(2\,\epsilon_1+\epsilon_2)\,\ln\left(\frac{k}{k_*}\right)\right]
\nonumber
\\
\label{P_SlowRoll}
\\
&&
\mathcal{P}_h=\frac{16\,H^2}{\pi\,m_{\rm Pl}^2}
\,\left[1-2\,(C+1)\,\epsilon_1-2\,\epsilon_1\,\ln\left(\frac{k}{k_*}
\right)\right]
\nonumber
\ ,
\ee
\end{widetext}
where $C\equiv\gamma_{_{\rm E}}+\ln 2-2\approx -\,0.7296$,
$\gamma_{_{\rm E}}$ being the Euler-Mascheroni constant,
and all quantities are evaluated at the Hubble crossing
$\eta_*$, that is the moment of time when
$k_*=(a\,H)(N_*)\equiv(a\,H)_*$.
(The number $k_*$ is usually called ``pivot scale''.)
From Eqs.~(\ref{P_SlowRoll}), we can also obtain the spectral
indices and $\alpha$-runnings,
\be
n_S-1=-2\,\epsilon_1-\epsilon_2 \ ,\quad n_T=-2\,\epsilon_1
\label{n_SlowRoll}
\ee
\be
\alpha_S=\alpha_T=0
\ ,
\label{alpha_SlowRoll}
\ee
on using respectively Eqs.~(\ref{n_def}) and (\ref{alpha_def}).
From Eq.~(\ref{R_def}), the tensor-to-scalar ratio becomes
\be
R=16\,\epsilon_1\left[1+C\,\epsilon_2+\epsilon_2
\,\ln\left(\frac{k}{k_*}\right)\right]
\ .
\label{R_SR}
\ee
\end{subequations}
\section{Leading WKB and second slow-roll order}
\label{1+2}
We now consider the possibility of obtaining consistent second
order results in the parameters $\epsilon_i$'s from the leading
WKB approximation.
\par
We first set $g_{(1)}^{\rm AD}(x)\equiv0$
in Eqs.~(\ref{spectra_correct}), (\ref{index_correct}),
(\ref{runnings_correct}) and (\ref{R_correct}).
We then find it convenient to re-express all relevant
quantities in terms of the conformal time $\eta$.
For this purpose, we employ the relation
\begin{subequations}
\be
-k\,\eta&=&
\frac{k}{a\,H\,(1-\epsilon_1)}\,
\left[1+\epsilon_1\,\epsilon_2+{\cal O}\left(\epsilon_i^3\right)\right]
\label{rel_change_var_2'ord}
\\
&=&
\frac{{\rm e}^{x}}{(1-\epsilon_1)}\,
\left[1+\epsilon_1\,\epsilon_2+{\cal O}\left(\epsilon_i^3\right)\right]
\ .
\label{change_var_2'ord}
\ee
\end{subequations}
We can now expand the frequencies to second order in the HFF,
\begin{widetext}
\be
\omega_{\rm S}^2(\eta)
&=&
k^2\,\eta^2\,\left(1-2\,\epsilon_1\,\epsilon_2\right)
-\left(\frac94+3\,\epsilon_1+4\,\epsilon_1^2 +\frac32\,\epsilon_
2+2\,\epsilon_1\,\epsilon_2
+\frac14\,\epsilon_2^2+\frac12\,\epsilon_2\,\epsilon_3\right)
\nonumber
\\
\label{omega2_2'ord}
\\
\omega _{\rm T}^2(\eta) &=& k^2\,\eta^2\,\left(1-2\,\epsilon_1\,\epsilon_2\right)
-\left(\frac94+3\,\epsilon_1+4\,\epsilon_1^2 -\frac12\,\epsilon_1\,\epsilon_2\right)
\nonumber
\ ,
\ee
and the PS become
\be
{\cal P}_\zeta
&=&\frac{H^2_{\rm f}}{\pi\,\epsilon_{1,{\rm f}}\,m_{\rm Pl}^2}\,
\left(-k\,\eta_{\rm f}\right)^3\,
\frac23\,\left(1-\frac83\,\epsilon_{1,{\rm f}}
-\frac13\,\epsilon_{2,{\rm f}}+\frac{19}{9}\,\epsilon_{1,{\rm f}}^2
-\frac{19}{9}\,\epsilon_{1,{\rm f}}\,\epsilon_{2,{\rm f}}
+\frac19\,\epsilon_{2,{\rm f}}^2
-\frac19\,\epsilon_{2,{\rm f}}\,\epsilon_{3,{\rm f}}\right)\,
{\rm e}^{2\xi_{\rm II, S}}
\nonumber
\\
\label{partial_spectra_2'ord}
\\
{\cal P}_h
&=&\frac{16\,H_{\rm f}^2}{\pi\,m_{\rm Pl}^2}\,
\left(-k\,\eta_{\rm f}\right)^3\,
\frac23\,\left(1-\frac83\,\epsilon_{1,{\rm f}}
+\frac{19}{9}\,\epsilon_{1,{\rm f}}^2
-\frac{26}{9}\,\epsilon_{1,{\rm f}}\,\epsilon_{2,{\rm f}}\right)\,
{\rm e}^{2\xi_{\rm II, T}}
\ .
\nonumber
\ee
\end{widetext}
From now on, the subscript $f$ will denote that the given quantity
is evaluated in the super-horizon limit.
\par
We must now compute the arguments of the exponentials
in the above expressions, which are in general of the form
\be
\xi_{\rm II}(\eta_0,\eta_{\rm f};k)
=\int_{\eta_{\rm f}}^{\eta_0}
\sqrt{A^2(\eta)-k^2\,\eta^2}\,\frac{\d\,\eta}{\eta}
\ ,
\label{int_with_change_2'ord}
\ee
where the function $A^2(\eta)$ contains the HFF,
but does not depend on $k$, and we have used
\be
\d\,x=\frac{\d\,\eta}{\eta}\,
\left[1+\epsilon_1(\eta)\,\epsilon_2(\eta)
+{\cal O}(\epsilon_i^3)\right]
\ .
\label{mesure_2'ord}
\ee
Let us also note that, at the turning point $\eta=\eta_0$,
one has
\be
A(\eta_0)=-k\,\eta_0
\ .
\label{TP}
\ee
It is now clear that, in order to obtain consistent results to
second order in the slow-roll parameters, we must consider
the time-dependence of the $\epsilon_i$'s, and the function
$A^2(\eta)$ may not be approximated by a constant.
This does not allow us to perform the integral unless the
scale factor $a=a(\eta)$ is given explicitly or, as we shall
see, some further approximation is employed.
\par
In Ref.~\cite{WKB_PRL}, we proposed a procedure which will
now be described in detail.
Let us start with the general exact relation
\be
&&
\!\!\!\!\!\!\!\!\!\!\!
\int_{\eta_1}^{\eta_2}\,
\sqrt{A^2(\eta)-k^2\,\eta^2}\,\frac{\d\,\eta}{\eta}
=
\left.
\sqrt{A^2(\eta)-k^2\,\eta^2}
\right|_{\eta_1}^{\eta_2}
\nonumber
\\
&&
\!\!\!\!\!\!\!\!\!\!\!
+
\left.
\frac{A(\eta)}{2}
\,\ln\left(\frac{A(\eta)-\sqrt{A^2(\eta)-k^2\,\eta^2}}
{A(\eta)+\sqrt{A^2(\eta)-k^2\,\eta^2}}\right)
\right|_{\eta_1}^{\eta_2}
\nonumber
\\
&&
\!\!\!\!\!\!\!\!\!\!\!
-\int_{\eta_1}^{\eta_2}
\,\ln\left(\frac{A(\eta)-\sqrt{A^2(\eta)-k^2\,\eta^2}}
{A(\eta)+\sqrt{A^2(\eta)-k^2\,\eta^2}}\right)\,
\frac{\left[A^2(\eta)\right]'}{4\,A(\eta)}\,\d\,\eta
\label{xi_gen_exact}
\ ,
\ee
which holds for every function $A^2(\eta)$.
We can derive this relation by initially considering the
integral in Eq.~(\ref{int_with_change_2'ord}) with $A$ constant
(i.e.~independent of $\eta$, which is just power-law inflation).
In this case, the integration can be performed and yields the first
and second term in the r.h.s.~of Eq.~(\ref{xi_gen_exact})
with the function $A$ independent of $\eta$,
whereas the third term vanishes identically.
If we then reinstate the $\eta$-dependence of $A$ in the two
non-vanishing terms and differentiate them with respect to $\eta_2$,
we obtain the original integrand in Eq.~(\ref{int_with_change_2'ord})
plus another term, originating from $A^2(\eta)$.
We then obtain the result~(\ref{xi_gen_exact}) by (formally) integrating
the latter term and subtracting it from the previous two.
\par
One can also repeat the above procedure for the new integral in the
r.h.s.~of Eq.~(\ref{xi_gen_exact}).
To do this, we define
\be
\left[A^2(\eta)\right]'
\equiv-\frac{1}{\eta}\,B(\eta)
\label{A2'_general}
\ ,
\ee
and, after several simplifications, we can write the general relation
as
\be
&&\!\!\!\!\!\!\!\!\!\!\!\!
-\int_{\eta_1}^{\eta_2}
\ln\left(\frac{A(\eta)-\sqrt{A^2(\eta)-k^2\,\eta^2}}
{A(\eta)+\sqrt{A^2(\eta)-k^2\,\eta^2}}\right)\,
\frac{\left[A^2(\eta)\right]'}{4\,A(\eta)}\,\d\,\eta
\nonumber
\\
&&\!\!\!\!\!\!\!\!\!\!\!\!
=
\left.
Y(\eta)\,\frac{B(\eta)}{8\,A(\eta)}
\right|_{\eta_1}^{\eta_2}
-\int_{\eta_1}^{\eta_2}
Y(\eta)\,
\left(\frac{B^2(\eta)}{16\,\eta\,A^3(\eta)}
+\frac{B'(\eta)}{8\,A(\eta)}\right)\,\d\eta
\nonumber
\\
&&
\!\!\!\!\!\!
-\int_{\eta_1}^{\eta_2}
\ln\left(\frac{A(\eta)-\sqrt{A^2(\eta)-k^2\,\eta^2}}
{A(\eta)+\sqrt{A^2(\eta)-k^2\,\eta^2}}\right)\,
\frac{B^2(\eta)}{8\,\eta\,A^3(\eta)}\,\d\eta
\label{last_int}
\ ,
\ee
with
\be
Y(\eta)&\equiv&
\ln\left(\frac{-k\,\eta}{2\,A(\eta)}\right)\,
\ln\left(\frac{A(\eta)-\sqrt{A^2(\eta)-k^2\,\eta^2}}
{A(\eta)+\sqrt{A^2(\eta)-k^2\,\eta^2}}\right)
\nonumber
\\
&&
+{\rm Li}_2\left(\frac{A(\eta)-\sqrt{A^2(\eta)-k^2\,\eta^2}}
{2\,A(\eta)}\right)
\nonumber
\\
&&
-{\rm Li}_2\left(\frac{A(\eta)+\sqrt{A^2(\eta)-k^2\,\eta^2}}
{2\,A(\eta)}\right)
\label{def_Y}
\ ,
\ee
and ${\rm Li}_2(z)$ is the dilogarithm function
(see, e.g.~Ref.~\cite{lewin}),
\be
{\rm Li}_2(z)\equiv\sum^\infty_{k=1}\,\frac{z^k}{k^2}
=-\int_0^z\,\frac{\ln\left(1-z'\right)}{z'}\,\d z'
\ .
\label{Li_def}
\ee
\par
On putting together all the pieces, we finally obtain
\begin{widetext}
\be
\int_{\eta_1}^{\eta_2}
\sqrt{A^2(\eta)-k^2\,\eta^2}\,\frac{\d\,\eta}{\eta}
&\!\!=\!\!&
\left.
\sqrt{A^2(\eta)-k^2\,\eta^2}
\right|_{\eta_1}^{\eta_2}
+
\left.
Y(\eta)\,\frac{B(\eta)}{8\,A(\eta)}
\right|_{\eta_1}^{\eta_2}
+
\left.
\frac{A(\eta)}{2}
\,\ln\left(\frac{A(\eta)-\sqrt{A^2(\eta)-k^2\,\eta^2}}
{A(\eta)+\sqrt{A^2(\eta)-k^2\,\eta^2}}\right)
\right|_{\eta_1}^{\eta_2}
\nonumber
\\
&&
-\int_{\eta_1}^{\eta_2}
Y(\eta)\,\left(\frac{B^2(\eta)}{16\,\eta\,A^3(\eta)}
+\frac{B'(\eta)}{8\,A(\eta)}\right)\,\d\eta
\nonumber
\\
&&
-\int_{\eta_1}^{\eta_2}
\ln\left(\frac{A(\eta)-\sqrt{A^2(\eta)-k^2\,\eta^2}}
{A(\eta)+\sqrt{A^2(\eta)-k^2\,\eta^2}}\right)\,
\frac{B^2(\eta)}{8\,\eta\,A^3(\eta)}\,\d\eta
\label{xi_gen_exact_compl}
\ .
\ee
\end{widetext}
\par
For the cases of interest to us [i.e.~for the
frequencies~(\ref{omega2_2'ord})], the functions $A$
for scalar and tensor perturbations to second order
in the HFF are given by
\begin{subequations}
\be
A^2_{\rm S}&=&
\frac94+3\,\epsilon_1+\frac32\,\epsilon_ 2
+4\,\epsilon_1^2+\frac{13}{2}\,\epsilon_1\,\epsilon_2
+\frac14\,\epsilon_2^2+\frac12\,\epsilon_2\,\epsilon_3
\nonumber
\\
\label{A2_2'ord}
\\
A^2_{\rm T}&=&
\frac94+3\,\epsilon_1+4\,\epsilon_1^2
+4\,\epsilon_1\,\epsilon_2
\nonumber
\ ,
\ee
so that
\be
B_{\rm S}=
3\,\epsilon_1\,\epsilon_ 2+\frac32\,\epsilon_ 2\,\epsilon_ 3
\ ,\quad
B_{\rm T}=
3\,\epsilon_1\,\epsilon_ 2
\label{B_2'ord}
\ .
\ee
\end{subequations}
For the results to second order in the HFF, we can
neglect the last two integrals in Eq.~(\ref{xi_gen_exact_compl}),
since $B'(\eta)=\mathcal{O}\left(\epsilon_i^3\right)$
and $B^2(\eta)=\mathcal{O}\left(\epsilon_i^4\right)$.
Thus our approximation consists in neglecting such terms and
not in assuming the HFF be constant (or any specific functional
form for them).
We can finally write $\xi_{\rm II, S}(\eta_0,\eta_{\rm f};k)$
and $\xi_{\rm II, T}(\eta_0,\eta_{\rm f};k)$ as
\begin{widetext}
\begin{subequations}
\be
\xi_{\rm II, S}
&\!\simeq\!&
-\frac32
+\frac{3\,\ln 3}{2}
+\ln 3\,\epsilon_{1,{\rm f}}
+\frac{\ln 3}{2}\,\epsilon_{2,{\rm f}}
+\left(\ln 3+\frac13\right)\,\epsilon_{1,{\rm f}}^2
+\frac{1}{12}\,\epsilon_{2,{\rm f}}^2
\nonumber
\\
&\!\!\!\!\!\!&
+\left(\frac{11\,\ln 3}{6}-\frac{\ln^2 3}{2}
+\frac{\pi^2}{24}+\frac13\right)\,\epsilon_{1,{\rm f}}\,\epsilon_{2,{\rm f}}
+\left(\frac{\ln 3}{6}-\frac{\ln^2 3}{4}+\frac{\pi^2}{48}\right)\,
\epsilon_{2,{\rm f}}\,\epsilon_{3,{\rm f}}
\nonumber
\\
&\!\!\!\!\!\!&
+\left[
-\frac32-\epsilon_{1,{\rm f}}-\frac12\,\epsilon_{2,{\rm f}}-\epsilon_{1,{\rm f}}^2
+\left(\ln 3-\frac{11}{6}\right)\,\epsilon_{1,{\rm f}}\,\epsilon_{2,{\rm f}}
+\left(\frac{\ln 3}{2}-\frac16\right)\,\epsilon_{2,{\rm f}}\,\epsilon_{3,{\rm f}}
\right]\,
\ln\left(-k\,\eta_{\rm f}\right)
\nonumber
\\
&\!\!\!\!\!\!&
-\frac12\,\left(\epsilon_{1,{\rm f}}\,\epsilon_{2,{\rm f}}
+\frac12\,\epsilon_{2,{\rm f}}\,\epsilon_{3,{\rm f}}\right)
\ln^2\left(-k\,\eta_{\rm f}\right)
\label{xi_2'_order_with_eps_S}
\ee
\be
\xi_{\rm II, T}
&\!\simeq\!&
-\frac32
+\frac{3\,\ln 3}{2}
+\ln 3\,\epsilon_{1,{\rm f}}
+\left(\ln 3+\frac13\right)\,\epsilon_{1,{\rm f}}^2
+\left(\frac{4\,\ln 3}{3}-\frac{\ln^2 3}{2}+\frac{\pi^2}{24}\right)\,
\epsilon_{1,{\rm f}}\,\epsilon_{2,{\rm f}}
\nonumber
\\
&\!\!\!\!\!\!&
+\left[
-\frac32-\epsilon_{1,{\rm f}}-\epsilon_{1,{\rm f}}^2
+\left(\ln 3-\frac43\right)\,\epsilon_{1,{\rm f}}\,\epsilon_{2,{\rm f}}
\right]\,
\ln\left(-k\,\eta_{\rm f}\right)
-\frac12\,\epsilon_{1,{\rm f}}\,\epsilon_{2,{\rm f}}
\ln^2\left(-k\,\eta_{\rm f}\right)
\ ,
\label{xi_2'_order_with_eps_T}
\ee
\end{subequations}
\end{widetext}
where the arguments in $\xi_{\rm II}$ have been omitted.
On inserting the above quantities in Eqs.~(\ref{partial_spectra_2'ord}),
we can calculate the PS.
\par
In order to compare the PS from Eqs.~(\ref{partial_spectra_2'ord}),
(\ref{xi_2'_order_with_eps_S}) and (\ref{xi_2'_order_with_eps_T}),
with the slow-roll expressions in Eqs.~(\ref{P_SlowRoll}),
we need a relation between $H(N_{\rm f})$ and
$H(N_*)$, $\epsilon_1(N_{\rm f})$ and $\epsilon_1(N_*)$,
and so on, to second order in the HFF (hereafter, quantities
without a subscript will be evaluated at the Hubble crossing
$\eta_*$ corresponding to the pivot scale $k_*$).
We expand the parameters $\epsilon_{i,{\rm f}}$ to first order
in $\Delta N\equiv N_f-N_*$ in the numerators and to second
order in the denominator of the scalar spectrum,
\be
\frac{\epsilon_i(N_{\rm f})}{\epsilon_i}
\simeq
1+\epsilon_{i+1} \Delta N
+\frac12\,
\left(\epsilon_{i+1}^2+\epsilon_{i+1}\,\epsilon_{i+2}\right)
{\Delta N}^2
\ ,
\label{epsilon_n_expans}
\ee
and, analogously, we find
\be
\frac{H^2(N_{\rm f})}{H^2}
\simeq
1-2\,\epsilon_1\,\Delta N-
\left(\epsilon_1\,\epsilon_2-2\,\epsilon_1^2\right)\,
{\Delta N}^2
\ .
\label{Hquad_expans_2'ord}
\ee
We can eliminate $\eta_{\rm f}$ in the logarithms
by expressing it in terms of $1/(a\,H)_{\rm f}$,
\be
\ln(-k\,\eta_{\rm f})
&\simeq&
\ln\left[\frac{k\,\left(1+\epsilon_{1,{\rm f}}\,
\epsilon_{2,{\rm f}}\right)}
{(a\,H)_{\rm f}\,\left(1-\epsilon_{1,{\rm f}}\right)}
\,\frac{(a\,H)_*}{(a\,H)_*}\right]
\nonumber
\\
&\simeq&
\ln\left(\frac{k}{k_*}\right)
-\Delta N
+\epsilon_{1,{\rm f}}\,\Delta N
+\epsilon_{1,{\rm f}}
\nonumber
\\
&\simeq&
\ln\left(\frac{k}{k_*}\right)
-\Delta N
+\epsilon_1\,\Delta N
+\epsilon_1
\ ,
\label{ln_expans_2'ord}
\ee
where in the first equality we used Eq.~(\ref{rel_change_var_2'ord}),
in the second one the definition of the pivot scale, and,
in the last relation, Eq.~(\ref{epsilon_n_expans}).
If we now use the above expressions we obtain the following PS,
where the superscript $^{(2)}$ stands for second slow-roll order
and the subscript WKB for leading adiabatic order,
\begin{subequations}
\begin{widetext}
\be
\mathcal{P}_{\zeta,\scriptscriptstyle\scriptscriptstyle{\rm WKB}}^{(2)}
&\!\!\!=\!\!\!&
\frac{H^2}{\pi\,\epsilon_1\,m_{\rm Pl}^2}\,
A_{\scriptscriptstyle{\rm WKB}}
\left\{1-2\left(D_{\scriptscriptstyle{\rm WKB}}+1\right)\,\epsilon_1
-D_{\scriptscriptstyle{\rm WKB}}\,\epsilon_2
+\left(2\,D_{\scriptscriptstyle{\rm WKB}}^2
+2\,D_{\scriptscriptstyle{\rm WKB}}-\frac19\right)\,\epsilon_1^2
\right.
\nonumber
\\
&&\left.
+\left(D_{\scriptscriptstyle{\rm WKB}}^2-D_{\scriptscriptstyle{\rm WKB}}
+\frac{\pi^2}{12}-\frac{20}{9}\right)\,\epsilon_1\,\epsilon_2
+\left(\frac12\,D_{\scriptscriptstyle{\rm WKB}}^2+\frac29\right)\,\epsilon_2^2
+\left(-\frac12\,D_{\scriptscriptstyle{\rm WKB}}^2+\frac{\pi^2}{24}-
\frac{1}{18}\right)\,\epsilon_2\,\epsilon_3
\right.
\nonumber
\\
&&\left.
+\left[-2\,\epsilon_1-\epsilon_2
+2\left(2\,D_{\scriptscriptstyle{\rm WKB}}+1\right)\,\epsilon_1^2
+\left(2\,D_{\scriptscriptstyle{\rm WKB}}-1\right)\,\epsilon_1\,\epsilon_2
+D_{\scriptscriptstyle{\rm WKB}}\,\epsilon_2^2
-D_{\scriptscriptstyle{\rm WKB}}\,\epsilon_2\,\epsilon_3\right]\,
\ln\left(\frac{k}{k_*}\right)
\right.
\nonumber
\\
&&
\left.
+\frac12\,\left(4\,\epsilon_1^2+2\,\epsilon_1\,\epsilon_2+\epsilon_2^2
-\epsilon_2\,\epsilon_3\right)\,
\ln^2\left(\frac{k}{k_*}\right)\right\}
\nonumber
\\
\label{P_SlowRoll_0_2order}
\\
\mathcal{P}_{h,\scriptscriptstyle{\rm WKB}}^{(2)}
&\!\!\!=\!\!\!&\frac{16\,H^2}{\pi\,m_{\rm Pl}^2}\,
A_{\scriptscriptstyle{\rm WKB}}
\left\{1-2\left(D_{\scriptscriptstyle{\rm WKB}}+1\right)\,\epsilon_1
+\left(2\,D_{\scriptscriptstyle{\rm WKB}}^2
+2\,D_{\scriptscriptstyle{\rm WKB}}-\frac19\right)\,\epsilon_1^2
\nonumber
\right.
\\
&&
\left.
+\left(-D_{\scriptscriptstyle{\rm WKB}}^2
-2\,D_{\scriptscriptstyle{\rm WKB}}+\frac{\pi^2}{12}-\frac{19}{9}\right)\,
\epsilon_1\,\epsilon_2
+\left[-2\,\epsilon_1+2\left(2\,D_{\scriptscriptstyle{\rm WKB}}+1\right)\,
\epsilon_1^2
-2\,\left(D_{\scriptscriptstyle{\rm WKB}}+1\right)\epsilon_1\,
\epsilon_2\right]\,
\ln\left(\frac{k}{k_*}\right)
\nonumber
\right.
\\
&&
\left.
+\frac12\,\left(4\,\epsilon_1^2
-2\,\epsilon_1\,\epsilon_2\right)\,\ln^2\left(\frac{k}{k_*}\right)\right\}
\ .
\nonumber
\ee
\end{widetext}
In the above, we have defined
$D_{\scriptscriptstyle{\rm WKB}}\equiv\frac{1}{3}-\ln 3\approx -0.7653$,
which approximates the coefficient $C$ in Eqs.~(\ref{P_SlowRoll}) with an
error of about $5\%$, and the factor
$A_{\scriptscriptstyle{\rm WKB}}\equiv 18/e^3\approx0.896$ which
gives an error of about $10\%$ on the amplitudes.
One of the main results of this paper, already reported in Ref.~\cite{WKB_PRL},
is that the PS to second order in the slow-roll parameters do not
depend on $\Delta N$, once the laborious dependence on the Hubble
crossing has been worked out: this fact is in complete agreement with the
constancy of the growing modes of $\zeta$ and $h$ on large scales.
The spectral indices, from Eqs.~(\ref{index_correct}),
and their runnings, from Eqs.~(\ref{runnings_correct}),
are analogously given by
\begin{widetext}
\be
n_{\rm S,\scriptscriptstyle{\rm WKB}}^{(2)}-1&=&
-2\,\epsilon_1-\epsilon_2-2\,\epsilon_1^2
-\left(2\,D_{\scriptscriptstyle{\rm WKB}}+3\right)\,\epsilon_1\,\epsilon_2
-D_{\scriptscriptstyle{\rm WKB}}\,\epsilon_2\,\epsilon_3
-2\,\epsilon_1\,\epsilon_2\,\ln\left(\frac{k}{k_*}\right)
-\epsilon_2\,\epsilon_3\,\ln\left(\frac{k}{k_*}\right)
\nonumber
\\
\label{n_2'order}
\\
n_{\rm T,\scriptscriptstyle{\rm WKB}}^{(2)}&=&
-2\,\epsilon_1-2\,\epsilon_1^2
-2\,\left(D_{\scriptscriptstyle{\rm WKB}}+1\right)\,\epsilon_1\,\epsilon_2
-2\,\epsilon_1\,\epsilon_2\,\ln\left(\frac{k}{k_*}\right)
\nonumber
\ee
\be
\alpha_{\rm S,\scriptscriptstyle{\rm WKB}}^{(2)}=
-2\,\epsilon_1\,\epsilon_2
-\epsilon_2\,\epsilon_3
\ ,
\quad
\alpha_{\rm T,\scriptscriptstyle{\rm WKB}}^{(2)}=
-2\,\epsilon_1\,\epsilon_2
\label{alpha_2'order}
\ .
\ee
From Eq.~(\ref{R_correct}) the tensor-to-scalar ratio becomes
\be
R^{(2)}_{\scriptscriptstyle{\rm WKB}}&=&
16\,\epsilon_1\left[1+D_{\scriptscriptstyle{\rm WKB}}\,\epsilon_2
+\left(D_{\scriptscriptstyle{\rm WKB}}+\frac19\right)\,\epsilon_1\,\epsilon_2
+\left(\frac12\,D_{\scriptscriptstyle{\rm WKB}}^2-\frac{2}{9}\right)\,\epsilon_2^2
+\left(\frac12\,D_{\scriptscriptstyle{\rm WKB}}^2-\frac{\pi^2}{24}
+\frac{1}{18}\right)\,\epsilon_2\,\epsilon_3
\right.
\nonumber
\\
&&\left.
+\left(\epsilon_2+\epsilon_1\,\epsilon_2+D_{\scriptscriptstyle{\rm WKB}}\,\epsilon_2^2
+D_{\scriptscriptstyle{\rm WKB}}\,\epsilon_2\,\epsilon_3\right)\,
\ln\left(\frac{k}{k_*}\right)
+\frac12\,\left(\epsilon_2^2+\epsilon_2\,\epsilon_3\right)\,
\ln^2\left(\frac{k}{k_*}\right)\right]
\ .
\label{R_lead_WKB}
\ee
\end{widetext}
\end{subequations}
We note that Eqs.~(\ref{P_SlowRoll_0_2order}),
(\ref{n_2'order}) and (\ref{alpha_2'order})
agree with Ref.~\cite{MS_WKB}, to first order in the HFF.
\par
We can also compare the above results with others as we
show in Appendix~\ref{app_other_res}.
For example, one could approximate the HFF by their Taylor
expansions around $\eta_0$ as in Ref.~\cite{H_MP,HHHJM,HHHJ},
and give the results in terms of the $\epsilon_i$'s calculated
at that point (see Table~\ref{H_MP_tab_PRD71}
for results only on spectral indices).
\section{Next-to-leading WKB and first slow-roll order}
\label{2+1}
To first order in the HFF, the function $A$ can be taken as
constant.
In this case, let us then introduce the more convenient
notation
\be
\xi_\pm(x_0,x;k)
&=&\pm\,\int_{x}^{x_0}\,\sqrt{\mp\,\omega^2(y)}\,\d\,y
\nonumber
\\
&\simeq&\pm\,\int_{\eta}^{\eta_0}
\sqrt{\mp\left(k^2\,\tau^2-A^2\right)}\,\frac{\d\,\tau}{\tau}
\nonumber
\\
&=&\pm\,A\,
\left[{\rm atan}_\pm\left(\Theta_\pm\right)-\Theta_\pm\right]
\ ,
\label{xi_I}
\ee
where we have used $\d\,x=\d\,\tau/\tau$, and
\begin{subequations}
\be
A^2_{\rm S}(\epsilon_i)&=&
\frac94+3\epsilon_1+\frac32\,\epsilon_2
\nonumber
\\
\label{B_S}
\\
A^2_{\rm T}(\epsilon_i)&=&
\frac94+3\epsilon_1
\nonumber
\ee
\be
k\,\eta_0=-A
\label{eta*}
\ ,
\ee
\end{subequations}
with $\epsilon_1$ and $\epsilon_2$ constant as well.
It is also useful to define
\begin{subequations}
\be
\Theta_\pm \equiv \sqrt{\pm\left(1-\frac{\eta^2}{\eta_0^2}\right)}
\label{arg_xi_I}
\ee
\be
&&
{\rm atan}_+(x) \equiv {\rm arctanh}(x)
\nonumber
\\
\\
&&
{\rm atan}_-(x) \equiv \arctan(x)
\ ,
\nonumber
\ee
\end{subequations}
where the plus (minus) sign corresponds to region~II (I).
With this notation, the expressions for $\gamma_{(1)}(\eta)$
and $\phi_{(1)}(\eta)$ in Eqs.~(\ref{phi_1})
can be explicitly written as
\begin{widetext}
\be
&
\gamma_{\pm(1)}(\eta)=\strut\displaystyle
\frac{1}{A^2}\,\left\{
-\frac{5\mp3\,\Theta_\pm^2}{24\,\Theta_\pm^4}
\pm\frac{5}{72\,\Theta_\pm\,
\left[{\rm atan}_\pm\left(\Theta_\pm\right)-\Theta_\pm\right]}
\right\}
&
\nonumber
\\
\\
&
\phi_{\pm(1)}(\eta)=\strut\displaystyle
\frac{1}{A^2}\,\left\{
\frac{23}{3150}
\pm\frac{505\mp654\,\Theta_\pm^2+153\,\Theta_\pm^4}{1152\,\Theta_\pm^6}
-\frac{5\,\left[\left(102\mp90\,\Theta_\pm^2\right)\,{\rm atan}_\pm
\left(\Theta_\pm\right)
-\left(102\mp157\,\Theta_\pm^2\right)\,\Theta_\pm\right]}
{10368\,\Theta_\pm^3\,\left[{\rm atan}_\pm\left(\Theta_\pm\right)
-\,\Theta_\pm\right]^2}
\right\}
\ .
&
\nonumber
\ee
\end{widetext}
The limits of interest are then given by
\begin{subequations}
\be
&&
\lim_{\eta\to-\infty}\,\gamma_{-(1)}(\eta)=0
\label{limit_gamma_I}
\\
&&
\lim_{\eta\to-\infty}\,\phi_{-(1)}(\eta)
=\frac{23}{3150\,A^2}
\label{limit_phi_I}
\\
&&
\lim_{\eta\to0^-}\,\gamma_{+(1)}(\eta)
=-\,\frac{1}{12\,A^2}
\label{limit_gamma_II}
\\
&&
\lim_{\eta\to0^-}\,\phi_{+(1)}(\eta)
=\frac{181}{16800\,A^2}
\ ,
\label{limit_phi_II}
\ee
where we have used
\be
\lim_{\eta\to-\infty}\,\Theta_-(\eta)=\infty
\label{limit_Theta-}
\\
\nonumber
\\
\lim_{\eta\to0^-}\,\Theta_+(\eta)=1
\ ,
\label{limit_Theta+}
\ee
and
\be
\lim_{\eta\to0^-}\,\left(
\omega_{\rm II}+\frac{\omega_{\rm II}'}{2\,\omega_{\rm II}}
\right)
=A
\ .
\label{limit_fun_in_spectra}
\ee
\end{subequations}
We can now use the values in Eqs.~(\ref{B_S})
to calculate the correction $g_{(1)}^{\rm AD}$ in the
super-horizon limit ($\eta\to 0^-$) for scalar
and tensor perturbations, obtaining
\be
&&
g_{(1){\rm S}}^{\rm AD}
=\frac{37}{324}
-\frac{19}{243}\,
\left(\epsilon_1+\frac12\,\epsilon_2\right)
\nonumber
\\
\label{AD_corrections}
\\
&&
g_{(1){\rm T}}^{\rm AD}
=\frac{37}{324}-\frac{19}{243}\,\epsilon_1
\ .
\nonumber
\ee
Finally, from Eqs.~(\ref{spectra_correct}) and (\ref{AD_corrections}),
we can write the expressions for the scalar and tensor PS,
with the superscript $^{(1)}$ for first slow-roll order and
the subscript WKB$*$ for next-to-leading adiabatic order,
as
\begin{subequations}
\begin{widetext}
\be
&&
\mathcal{P}_{\zeta,\scriptscriptstyle{\rm WKB*}}^{(1)}=
\frac{H^2}{\pi\,\epsilon_1\,m_{\rm Pl}^2}\,
A_{\scriptscriptstyle{\rm WKB*}}\,
\left[1-2\,\left(D_{\scriptscriptstyle{\rm WKB*}}+1\right)\,\epsilon_1
-D_{\scriptscriptstyle{\rm WKB*}}\,\epsilon_2
-\left(2\,\epsilon_1+\epsilon_2\right)\,\ln\left(\frac{k}{k_*}\right)\right]
\nonumber
\\
\label{P_SlowRoll_1}
\\
&&
\mathcal{P}_{h,\scriptscriptstyle{\rm WKB*}}^{(1)}=
\frac{16\,H^2}{\pi\,m_{\rm Pl}^2}\,
A_{\scriptscriptstyle{\rm WKB*}}\,
\left[1-2\,\left(D_{\scriptscriptstyle{\rm WKB*}}+1\right)\,\epsilon_1
-2\,\epsilon_1\,\ln\left(\frac{k}{k_*}\right)\right]
\ ,
\nonumber
\ee
\end{widetext}
where
$D_{\scriptscriptstyle{\rm WKB*}}\equiv\frac{7}{19}-\ln 3\approx -0.7302$,
which approximates the coefficient $C$ in Eqs.~(\ref{P_SlowRoll}) with
an error of about $0.08\%$, and the factor
$A_{\scriptscriptstyle{\rm WKB*}}\equiv 361/18\,e^3\approx 0.999$
which gives an error of about $0.1\%$ on the amplitudes.
We also obtain the standard slow-roll spectral indices
and $\alpha$-runnings,
\be
\!\!\!\!\!\!\!\!
n_{\rm S,\scriptscriptstyle{\rm WKB*}}^{(1)}-1=
-2\,\epsilon_1-\epsilon_2
\ ,\quad
n_{\rm T,\scriptscriptstyle{\rm WKB*}}^{(1)}=-2\,\epsilon_1
\label{n_NTL_WKB}
\ee
\be
\!\!\!\!\!\!\!\!
\alpha_{\rm S,\scriptscriptstyle{\rm WKB*}}^{(1)}=
\alpha_{\rm T,\scriptscriptstyle{\rm WKB*}}^{(1)}=0
\ ,
\label{alpha_NTL_WKB}
\ee
on respectively using Eqs.~(\ref{index_correct}) and
(\ref{runnings_correct}).
From Eq.~(\ref{R_correct}) the tensor-to-scalar ratio becomes
\be
R^{(1)}_{\scriptscriptstyle{\rm WKB*}}=
16\,\epsilon_1\left[1+D_{\scriptscriptstyle{\rm WKB*}}\,\epsilon_2
+\epsilon_2\,\ln\left(\frac{k}{k_*}\right)\right]
\ .
\label{R_next-to-lead_WKB}
\ee
\end{subequations}
\section{Next-to-leading WKB and second slow-roll order}
\label{2+2}
In order to give the results to next-to-leading WKB order
and second slow-roll order, we should evaluate the corrections
$g_{(1)}^{\rm AD}$ for scalar and tensor perturbations
to second order in the $\epsilon_i$'s.
This, in turn, would require the computation of the functions
$\phi_{\pm(1)}$ and $\gamma_{\pm(1)}$, in Eqs.~(\ref{phi_1}),
whose analytical calculation is extremely difficult.
We shall therefore try to give the expressions for the corrections
by following a heuristic (and faster) method.
\par
Let us start from the first order slow-roll expressions given in
Eqs.~(\ref{AD_corrections}) and add all possible second order terms
in the HFF.
Some of such terms can then be (partially) fixed by making reasonable
requirements, which will be explained below, so that the corrections
to the PS read
\begin{widetext}
\be
1+g_{(1){\rm S}}^{\rm AD}
&\!\!=\!\!&
\frac{361}{324}\,
\left\{1-\frac{4}{57}\,
\left(\epsilon_{1,{\rm f}}+\frac12\,\epsilon_{2,{\rm f}}\right)
-\frac{8}{361}\,\epsilon_{1,{\rm f}}^2
+\left[\frac{4}{57}\,\left({b}_{\rm S}-D_{\scriptscriptstyle{\rm WKB*}}\right)
-\frac{262}{3249}\right]\,\epsilon_{1,{\rm f}}\,\epsilon_{2,{\rm f}}
+\frac{13}{1083}\,\epsilon_{2,{\rm f}}^2
\right.
\nonumber
\\
&&
\phantom{\frac{361}{324}\,}
\left.
+\left[\frac{2}{57}\,\left({d}_{\rm S}-D_{\scriptscriptstyle{\rm WKB*}}\right)
-\frac{2}{171}\right]\,\epsilon_{2,{\rm f}}\,\epsilon_{3,{\rm f}}
-\left(\frac{4}{57}\,\epsilon_{1,{\rm f}}\,\epsilon_{2,{\rm f}}
+\frac{2}{57}\,\epsilon_{2,{\rm f}}\,\epsilon_{3,{\rm f}}\right)
\,\ln\left(-k\,\eta_{\rm f}\right)\right\}
\nonumber
\\
\label{AD_corrections_2'order}
\\
1+g_{(1){\rm T}}^{\rm AD}
&\!\!=\!\!&
\frac{361}{324}\,
\left\{1-\frac{4}{57}\,\epsilon_{1,{\rm f}}
-\frac{8}{361}\,\epsilon_{1,{\rm f}}^2
+\left[\frac{4}{57}\,\left({b}_{\rm T}-D_{\scriptscriptstyle{\rm WKB*}}\right)
-\frac{16}{171}\right]\,\epsilon_{1,{\rm f}}\,\epsilon_{2,{\rm f}}
-\frac{4}{57}\,\epsilon_{1,{\rm f}}\,\epsilon_{2,{\rm f}}
\,\ln\left(-k\,\eta_{\rm f}\right)\right\}
\ .
\nonumber
\ee
\end{widetext}
First of all, we have factorized the number $361/324$ so as
to recover the standard slow-roll amplitudes (\ref{P_SlowRoll})
with a very good accuracy.
As already mentioned, the first order terms are the same as those
in the results (\ref{AD_corrections}) of Section~\ref{2+1},
evaluated however in the super-horizon limit.
The form of the coefficients multiplying the second order monomials
in the HFF have been partially determined by using the expressions given in
Section~\ref{2+1} [Eq.~(\ref{limit_gamma_I})-(\ref{limit_fun_in_spectra})]
with Eqs.~(\ref{A2_2'ord}) replacing Eqs.~(\ref{B_S}).
By this procedure, one expects to obtain results correct up to
derivatives of $A^2$, which can just contain mixed terms to second order.
The numerical coefficients in front of $\epsilon_{1,{\rm f}}^2$ and
$\epsilon_{2,{\rm f}}^2$ are thus uniquely determined,
whereas the coefficients multiplying the mixed terms
$\epsilon_{1,{\rm f}}\,\epsilon_{2,{\rm f}}$ and
$\epsilon_{2,{\rm f}}\,\epsilon_{3,{\rm f}}$ remain partially
ambiguous.
The last number in each square bracket arises directly from
this procedure, whereas the other terms are chosen so as to match
the ``standard'' dependence on $D_{\scriptscriptstyle{\rm WKB*}}$
in the PS, with $b_{\rm S,T}$ and $d_{\rm S}$ left undetermined.
Finally, we have also required that the PS do not depend on
$\Delta N$, consistently with the constancy in time of the growing
modes for $h$ and $\zeta$.
This requirement fixes the coefficients in front of the logarithms in
Eqs.~(\ref{AD_corrections_2'order}) uniquely~\footnote{For the same
reason (i.e.~to avoid the appearance of $\Delta N$), we cannot consider
terms such as
$\mathcal{O}\,(\epsilon^2_i)\,\ln^2\left(-k\,\eta_{\rm f}\right)$
in Eqs.~(\ref{AD_corrections_2'order}).}, and leads to spectral indices
which do not depend on $b_{\rm S,T}$ and $d_{\rm S}$.
\par
Proceeding then as in Section~\ref{1+2}, from the above corrections
we obtain the expressions for the scalar and tensor PS,
\begin{widetext}
\begin{subequations}
\be
\mathcal{P}_{\zeta,\scriptscriptstyle{\rm WKB*}}^{(2)}&\!\!\!=\!\!\!&
\frac{H^2}{\pi\,\epsilon_1\,m_{\rm Pl}^2}\,
A_{\scriptscriptstyle{\rm WKB*}}\,
\left\{1-2\left(D_{\scriptscriptstyle{\rm WKB*}}+1\right)\,\epsilon_1
-D_{\scriptscriptstyle{\rm WKB*}}\,\epsilon_2
+\left(2\,{D^2_{\scriptscriptstyle{\rm WKB*}}}
+2\,D_{\scriptscriptstyle{\rm WKB*}}-\frac{71}{1083}\right)\,
\epsilon_1^2
\right.
\nonumber
\\
&&\left.
+\left({D^2_{\scriptscriptstyle{\rm WKB*}}}
-D_{\scriptscriptstyle{\rm WKB*}}+\frac{\pi^2}{12}
+{b}_{\rm S}\,\frac{4}{57}-\frac{2384}{1083}\right)\,
\epsilon_1\,\epsilon_2
+\left(\frac12\,{D^2_{\scriptscriptstyle{\rm WKB*}}}
+\frac{253}{1083}\right)\,\epsilon_2^2
\right.
\nonumber
\\
&&\left.
+\left(-\frac12\,{D^2_{\scriptscriptstyle{\rm WKB*}}}+\frac{\pi^2}{24}
+{d}_{\rm S}\,\frac{2}{57}-\frac{49}{722}\right)\,
\epsilon_2\,\epsilon_3
\right.
\nonumber
\\
&&\left.
+\left[-2\,\epsilon_1-\epsilon_2+2\left(2\,D_{\scriptscriptstyle{\rm WKB*}}+1\right)\,\epsilon_1^2
+\left(2\,D_{\scriptscriptstyle{\rm WKB*}}-1\right)\,\epsilon_1\,\epsilon_2
+D_{\scriptscriptstyle{\rm WKB*}}\,\epsilon_2^2
-D_{\scriptscriptstyle{\rm WKB*}}\,\epsilon_2\,\epsilon_3\right]\,
\ln\left(\frac{k}{k_*}\right)
\right.
\nonumber
\\
&&\left.
+\frac12\,\left(4\,\epsilon_1^2
+2\,\epsilon_1\,\epsilon_2+\epsilon_2^2
-\epsilon_2\,\epsilon_3\right)
\,\ln^2\left(\frac{k}{k_*}\right)\right\}
\nonumber
\\
\label{P_SlowRoll_1_2order}
\\
\mathcal{P}_{h,\scriptscriptstyle{\rm WKB*}}^{(2)}&\!\!\!=\!\!\!&
\frac{16\,H^2}{\pi\,m_{\rm Pl}^2}\,
A_{\scriptscriptstyle{\rm WKB*}}\,
\left\{
1-2\left(D_{\scriptscriptstyle{\rm WKB*}}+1\right)\,\epsilon_1
+\left(2\,{D^2_{\scriptscriptstyle{\rm WKB*}}}
+2\,D_{\scriptscriptstyle{\rm WKB*}}
-\frac{71}{1083}\right)\,\epsilon_1^2
\nonumber
\right.
\\
&&
\left.
+\left(-{D^2_{\scriptscriptstyle{\rm WKB*}}}
-2\,D_{\scriptscriptstyle{\rm WKB*}}+\frac{\pi^2}{12}
+{b}_{\rm T}\,\frac{4}{57}-\frac{771}{361}\right)\,\epsilon_1\,\epsilon_2
\nonumber
\right.
\\
&&
\left.
+\left[-2\,\epsilon_1+2\left(2\,D_{\scriptscriptstyle{\rm WKB*}}+1\right)\,\epsilon_1^2
-2\,\left(D_{\scriptscriptstyle{\rm WKB*}}+1\right)\epsilon_1\,\epsilon_2\right]
\,\ln\left(\frac{k}{k_*}\right)
+\frac12\,\left(4\,\epsilon_1^2
-2\,\epsilon_1\,\epsilon_2\right)\,\ln^2\left(\frac{k}{k_*}\right)\right\}
\ .
\nonumber
\ee
We also obtain the spectral indices~(\ref{index_correct})
and their runnings~(\ref{runnings_correct}),
\be
&&
n_{\rm S,\scriptscriptstyle{\rm WKB*}}^{(2)}-1=
-2\,\epsilon_1-\epsilon_2-2\,\epsilon_1^2
-\left(2\,D_{\scriptscriptstyle{\rm WKB*}}+3\right)\,\epsilon_1\,\epsilon_2
-D_{\scriptscriptstyle{\rm WKB*}}\,\epsilon_2\,\epsilon_3
-2\,\epsilon_1\,\epsilon_2\ln\left(\frac{k}{k_*}\right)
-\epsilon_2\,\epsilon_3\ln\left(\frac{k}{k_*}\right)
\nonumber
\\
\label{n_2'order_next_WKB}
\\
&&
n_{\rm T,\scriptscriptstyle{\rm WKB*}}^{(2)}=
-2\,\epsilon_1-2\,\epsilon_1^2
-2\,\left(D_{\scriptscriptstyle{\rm WKB*}}+1\right)\,\epsilon_1\,\epsilon_2
-2\,\epsilon_1\,\epsilon_2\ln\left(\frac{k}{k_*}\right)
\nonumber
\ee
\be
\alpha_{\rm S,\scriptscriptstyle{\rm WKB*}}^{(2)}=
-2\,\epsilon_1\,\epsilon_2
-\epsilon_2\,\epsilon_3
\ ,\quad
\alpha_{\rm T,\scriptscriptstyle{\rm WKB*}}^{(2)}=
-2\,\epsilon_1\,\epsilon_2
\label{alpha_2'order_next_WKB}
\ ,
\ee
and the tensor-to-scalar ratio becomes
\be
R_{\scriptscriptstyle{\rm WKB*}}^{(2)}&=&
16\,\epsilon_1\left\{1+D_{\scriptscriptstyle{\rm WKB*}}\,\epsilon_2
+\left[D_{\scriptscriptstyle{\rm WKB*}}
+\left({b}_{\rm T}-{b}_{\rm S}\right)\,\frac{4}{57}
+\frac{71}{1083}\right]\,\epsilon_1\,\epsilon_2
\right.
\nonumber
\\
&&\left.
+\left(\frac12\,{D^2_{\scriptscriptstyle{\rm WKB*}}}
-\frac{253}{1083}\right)\,\epsilon_2^2
+\left(\frac12\,{D^2_{\scriptscriptstyle{\rm WKB*}}}-\frac{\pi^2}{24}
-{d}_{\rm S}\,\frac{2}{57}+\frac{49}{722}\right)
\,\epsilon_2\,\epsilon_3
\right.
\nonumber
\\
&&\left.
+\left(\epsilon_2+\epsilon_1\,\epsilon_2
+D_{\scriptscriptstyle{\rm WKB*}}\,\epsilon_2^2
+D_{\scriptscriptstyle{\rm WKB*}}\,\epsilon_2\,\epsilon_3\right)\,
\ln\left(\frac{k}{k_*}\right)
+\frac12\,\left(\epsilon_2^2+\epsilon_2\,\epsilon_3\right)\,
\ln^2\left(\frac{k}{k_*}\right)\right\}
\ .
\label{R_next-to-lead_WKB_2'order}
\ee
\end{subequations}
\end{widetext}
\par
Let us end this Section with a few remarks.
The undetermined coefficients $b_{\rm S,T}$ and $d_{\rm S}$
still appear in the PS (and their ratio $R$), but not in the spectral
indices and runnings, which are therefore uniquely specified by our
(heuristic) procedure.
In general, one may expect that the complete adiabatic corrections
to second order contain HFF calculated both in the super-horizon
limit and at the turning points (zeros) of the frequencies.
However, this is not relevant in the present context, since the possible
numerical coefficients that would multiply such terms can be eventually
re-absorbed in the definitions of ${b}_{\rm S,T}$ and ${d}_{\rm S}$.
This could only be an issue (in particular for the spectral indices)
if we considered third order terms (since it would then become important where
the second order terms are evaluated).
\section{Conclusions}
\label{conc}
We have shown that the improved WKB treatment of cosmological
perturbations agrees with the standard slow-roll
approximation~\cite{SL} to within $0.1\,\%$, finally resolving the
issue of a $10\,\%$ error in the prediction of the amplitudes
to lowest order which was raised in Ref.~\cite{MS_WKB}.
\par
The next issues are the inflationary predictions to second order
in the slow-roll parameters.
After the results on the running of the power spectra~\cite{KT},
second order results have been obtained using the Green's function
method with the massless solution in a de~Sitter
space-time~\cite{SGong,LLMS}.
In a previous manuscript~\cite{WKB_PRL}, and in full detail here,
we have employed the leading WKB approximation to obtain the scalar
and tensor power spectra to second order in the slow-roll parameters.
The key technical point is to use Eq.~(\ref{xi_gen_exact_compl}),
which allows one to express the power spectra on large scales as a
function of the Hubble crossing quantities, but with no explicit
dependence on $\Delta N$.
As one of our main results, we find that the polynomial structure of
the power spectra in the $\epsilon_i$'s obtained from the WKB method
agrees with the one arising from the Green's function approach of
Refs.~\cite{SGong,LLMS}.
\par
In Section~\ref{2+2}, as a further development of Ref.~\cite{WKB_PRL},
we employ the next-to-leading WKB approximation to second order in the
$\epsilon_i$'s and, by requiring that the power spectra do not explicitly
depend on $\Delta N$ (a property which instead was derived both in
Sections~\ref{1+2} and~\ref{2+1}), and using the expressions found
in Section~\ref{2+1}, we obtain unique expressions for the spectral
indices and runnings.
Our findings show that different ways of approximating cosmological
perturbations show up at the leading order in the amplitude, but in the
next-to-leading order in the derivatives of the power-spectra with respect
to wave number $k$.
The accuracy in the theoretical predictions on spectral indices to second
order in the slow-roll parameters, evaluated at $k=k_*$, is now striking:
the Green's function method coefficient $C\simeq -0.7296$ is
replaced by $D_{\rm WKB*}\simeq -0.7302$ in the WKB method, leading to a
precision of $1$ part on $1000$ in the predictions of the coefficients
of ${\cal O}(\epsilon_i^2)$ terms in the spectral indexes.
\appendix
\section{Slow-Roll {\em vs} Horizon Flow}
\label{app}
In this Appendix, we compare the HFF we used in the text
with some other (equivalent) hierarchies.
\par
In Table~\ref{convers1}, we give the relations between the
slow-roll parameters defined by some other authors and the HFF
defined in Eq.~(\ref{hor_flo_fun}).
\begin{table}[ht]
\caption{Relation between (some) slow-roll parameters and the HFF
used here and in Ref.~\cite{MS_WKB}.
\label{convers1}}
\centerline{
\begin{tabular}{ c c }
\hline
\hline
\hspace{\stretch{1}} J.~Martin \textit{et al.} \cite{MS}$^{\rm a}$
\hspace{\stretch{1}}
& \hspace{\stretch{1}} E.D.~Stewart \textit{et al.} \cite{SL,SGong}$^{\rm b}$
\hspace{\stretch{1}}
\\
\hspace{\stretch{1}} S.~Habib \textit{et al.} \cite{H_MP}
\hspace{\stretch{1}}
& \hspace{\stretch{1}} S.~Habib \textit{et al.} \cite{HHHJM,HHHJ}
\hspace{\stretch{1}}
\\
\hline
$\epsilon=\epsilon_1$
& $\epsilon=\epsilon_1$
\\ 
$\delta=\epsilon_1-\frac{1}{2}\,\epsilon_2$
& $\delta_{1}=\frac{1}{2}\,\epsilon_2-\epsilon_1$
\\ 
$\xi_2=\frac{1}{2}\,\epsilon_2\,\epsilon_3$
& $\delta_2=-\frac{5}{2}\,\epsilon_1\,\epsilon_2+2\,\epsilon_1^2
+\frac{1}{4}\,\epsilon_2^2+\frac{1}{2}\,\epsilon_2\,\epsilon_3$
\\ 
\hline
\hline
\end{tabular}
}
$^{\rm a}$ In Ref.~\cite{MS} $\xi$ is used instead of
$\xi_2$.
$^{\rm b}$In Ref.~\cite{SL} $\epsilon_1$, $\delta$, and
${\dddot{\phi}\,\delta}/H\,\ddot{\phi}$ are used instead of $\epsilon$,
$\delta_1$ and $\delta_2$ respectively.
\end{table}
\par
We also compare the Hubble-slow-roll~(HSR) and the potential-slow-roll~(PSR)
parameters (see~Ref.~\cite{LPB}) with the HFF.
We recall that some authors redefine $\xi_V$ and $\xi_H$ as
$\xi_V^2$ and $\xi_H^2$ and write only $\epsilon$, $\eta$ and $\xi$,
which can be confusing.
The relations between the $\epsilon_i$'s and HSR parameters are
\be
&&
\epsilon_H=\epsilon_1
\ ,
\nonumber
\quad
\eta_H=\epsilon_1-\frac12\,\epsilon_2
\nonumber
\\
\label{HFFvsHSR}
\\
&&
\xi_H^2=\epsilon_1^2-\frac32\,\epsilon_1\,\epsilon_2
+\frac12\,\epsilon_2\,\epsilon_3
\nonumber
\ ,
\ee
and for the PSR parameters we have
\begin{widetext}
\be
\epsilon_V
&\!=\!&
\epsilon_1\,\left[1+\frac{\epsilon_2}{2\,(3-\epsilon_1)}\right]^2
\sim
\epsilon_1
\ ,
\quad
\eta_V\!=\!
\frac{6\,\epsilon_1-\frac32\,\epsilon_2-2\,\epsilon_1^2
+\frac52\,\epsilon_1\,\epsilon_2
-\frac{1}{4}\,\epsilon_2^2
-\frac12\,\epsilon_2\,\epsilon_3}{3-\epsilon_1}
\sim
2\,\epsilon_1-\frac12\,\epsilon_2
\nonumber
\\
\nonumber
\\
\xi_V^2\!&=&\!\frac{\left(2\,\epsilon_1-6-\epsilon_2\right)
\left[8\,\epsilon_1^3-6\,\epsilon_1^2\,(4+3\,\epsilon_2)
+\epsilon_1\,\epsilon_2\,(18+6\,\epsilon_2+7\,\epsilon_3)
-\epsilon_2\,\epsilon_3\,(3+\epsilon_2+\epsilon_3+\epsilon_4)\right]}
{4\,\left(3-\,\epsilon_1\right)^2}
\label{HFFvsPSR}
\\
&\sim&
4\,\epsilon_1^2
-3\,\epsilon_1\,\epsilon_2+\frac12\,\epsilon_2\,\epsilon_3
\ ,
\nonumber
\ee
where we have shown both the exact formulas and approximate
relations to the order of interest.
\par
We complete this review with the first three exact connection
formulae between the PSR and HSR parameters~\cite{LPB}
\be
&
\strut\displaystyle\frac{\epsilon_V}{\epsilon_H}=
\frac{\left(3-\eta_H\right)^2}{\left(3-\epsilon_H\right)^2}
\ ,
\quad
\eta_V=
\strut\displaystyle\frac{3\left(\epsilon_H+\eta_H\right)-\eta_H^2-\xi_H^2}
{3-\epsilon_H}
&
\nonumber
\\
\label{PSRvsHSR}
\\
&
\xi_V^2=
\strut\displaystyle\frac{9\,\left(3\epsilon_H\eta_H+\xi_H^2-\epsilon_H\eta_H^2\right)
-3\left(4\eta_H\xi_H^2-\eta_H^2\xi_H^2+\sigma_H^3\right)+\eta_H\sigma_H^3}
{(3-\epsilon_H)^2}
\ ,
&
\nonumber
\ee
and so on with $\sigma_H$ the next HSR parameter.
\end{widetext}
\section{Other results in the literature}
\label{app_other_res}
We use Table~\ref{convers1} and Eqs.~(\ref{HFFvsHSR}),
to compare the results for spectral indices
- and for PS and runnings when available - in terms
of the HFF in Tables~\ref{LLMS_SG_tab} and~\ref{H_MP_tab_PRD71}.
\par
\begin{table*}[!ht]
\caption{Results obtained with the Green's function method~\cite{LLMS}
suggested in Ref.~\cite{SGong};
$C\equiv\ln 2+\gamma_E-2\simeq  -0.7296$ with $\gamma_E$
the Euler-Mascheroni constant.
We display the original results for spectral indices,
$\alpha$-runnings and tensor-to-scalar ratio evaluated at the pivot
scale $k_*$.}
\centerline{
\begin{tabular}{ l }
\hline
\hline
\centerline{
E.D.~Stewart \textit{et al.} \cite{SL},
E.D.~Stewart \textit{et al.} \cite{SGong},
S.M.~Leach \textit{et al.} \cite{LLMS},
A.R.~Liddle \textit{et al.} \cite{LPB}.
}
\\
\hline
\\
$
\mathcal{P}_\zeta=
\frac{H^2}{\pi\,\epsilon_1\,m_{\rm Pl}^2}\,
\left\{
1
-2\left(C+1\right)\,
\epsilon_1
-C\,
\epsilon_2
+\left(2\,C^2+2\,C+\frac{\pi^2}{2}-5\right)\,
\epsilon_1^2
\right.$
\\
\\
$\phantom{aaaa}\left.
+\left(C^2-C+\frac{7\,\pi^2}{12}-7\right)\,
\epsilon_1\,\epsilon_2
+\left(\frac12\,C^2+\frac{\pi^2}{8}-1\right)\,
\epsilon_2^2
+\left(-\frac12\,C^2+\frac{\pi^2}{24}\right)\,
\epsilon_2\,\epsilon_3
\right.$
\\
\\
$\phantom{aaaa}\left.
+\left[-2\,\epsilon_1-\epsilon_2+2\left(2\,C+1\right)\,\epsilon_1^2
+\left(2\,C-1\right)\,\epsilon_1\,\epsilon_2
+C\,\epsilon_2^2
-C\,\epsilon_2\,\epsilon_3\right]\,
\ln\left(\frac{k}{k_*}\right)
+\frac12\,\left(4\,\epsilon_1^2
+2\,\epsilon_1\,\epsilon_2+\epsilon_2^2
-\epsilon_2\,\epsilon_3\right)
\,\ln^2\left(\frac{k}{k_*}\right)
\right\}$
\\
\\
$\mathcal{P}_h=
\frac{16\,H^2}{\pi\,m_{\rm Pl}^2}\,
\left\{
1-2\left(C+1\right)\,\epsilon_1
+\left(2\,C^2+2\,C+\frac{\pi^2}{2}-5\right)\,\epsilon_1^2
+\left(-C^2-2\,C+\frac{\pi^2}{12}-2\right)\,
\epsilon_1\,\epsilon_2
\nonumber
\right.$
\\
\\
$\phantom{aaaa}\left.
+\left[-2\,\epsilon_1+2\left(2\,C+1\right)\,\epsilon_1^2
-2\,\left(C+1\right)\epsilon_1\,\epsilon_2\right]
\,\ln\left(\frac{k}{k_*}\right)
+\frac12\,\left(4\,\epsilon_1^2
-2\,\epsilon_1\,\epsilon_2\right)\,\ln^2\left(\frac{k}{k_*}\right)
\right\}$
\\
\\
$n_{\rm S}-1=-2\,\epsilon_1-\epsilon_2-2\,\epsilon_1^2
-\left(2\,C+3\right)\,\epsilon_1\,\epsilon_2
-C\,\epsilon_2\,\epsilon_3$
\ ,\quad
$n_{\rm T}=-2\,\epsilon_1-2\,\epsilon_1^2-2\,\left(C+1\right)\,\epsilon_1\,\epsilon_2$
\\
\\
$\alpha_{\rm S}=-2\,\epsilon_1\,\epsilon_2-\epsilon_2\,\epsilon_3$
\ ,\quad
$\alpha_{\rm T}=-2\,\epsilon_1\,\epsilon_2$
\\
\\
$R=16\,\epsilon_1\left[1+C\,\epsilon_2
+\left(C-\frac{\pi^2}{2}+5\right)\,\epsilon_1\,\epsilon_2
+\left(\frac12\,C^2-\frac{\pi^2}{8}+1\right)\,\epsilon_2^2
+\left(\frac12\,C^2-\frac{\pi^2}{24}\right)\,\epsilon_2\,\epsilon_3
\right]$
\\
\\
\hline
\hline
\end{tabular}
}
\label{LLMS_SG_tab}
\end{table*}
In Table~\ref{LLMS_SG_tab} we summarize the results for scalar and tensor
perturbations given in Ref.~\cite{LLMS}, as calculated with the method
proposed in Ref.~\cite{SGong}.
We also refer to some other papers which give the same results,
although expressed in different hierarchies.
\par
Let us then compare with the results obtained by the uniform
approximation~\cite{H_MP,HHHJM,HHHJ}.
The integrand in Eq.~(9) of Ref.~\cite{H_MP} can be obtained
from our Eq.~(\ref{osci}) by the change of variables
$x=\ln\left(-k\,\eta\right)$, $u={\rm e}^{-x/2}\,\mu$, which
of course differs from our Eq.~(\ref{transf}).
Therefore, the integrands in our Eq.~(\ref{xiI_II}) and Eq.~(9) of
Ref.~\cite{H_MP} differ by ${\cal O}(\epsilon_i^4)$ terms, once
both are expressed in conformal time.
We also note that in Refs.~\cite{H_MP,HHHJM,HHHJ}, the authors just give
expressions for the spectral indices which depend on $k$ (i.e.~without
referring to a pivot scale $k=k_*$).
Further, it is not easy to see that they obtain the correct scaling in
$\ln\left(k/k_*\right)$, which yields the $\alpha$-runnings, and
the $\epsilon_i$'s are evaluated at the classical turning points for the
frequencies.
\par
\begin{table}[!h]
\caption{Results obtained with the uniform approximation.
\label{H_MP_tab_PRD71}}
\centerline{
\begin{tabular}{ c }
\hline
\hline
S.~Habib \textit{et al.}~\cite{HHHJ}
\\
\hline
$n_{\rm S}-1=-2\,\epsilon_1-\epsilon_2-2\,\epsilon_1^2
-\left(\frac{20}{3}-2\,\pi\right)\,\epsilon_1\,\epsilon_2
-\left(\frac{11}{6}-\pi\right)\,\epsilon_2\,\epsilon_3$
\\
$n_{\rm T}=-2\,\epsilon_1-2\,\epsilon_1^2
-\left(\frac{14}{3}-\frac{3}{2}\pi\right)\,\epsilon_1\,\epsilon_2$
\\
\hline
\hline
\end{tabular}
}
\end{table}
In Tables~\ref{H_MP_tab_PRD71}, we
display the results for scalar and tensor perturbations
given in Ref.~\cite{HHHJ}.
As it was properly stated in Ref.~\cite{HHHJ}, the results for scalar
and tensor spectral indices given in Ref.~\cite{H_MP,HHHJM} were not
fully expanded to second order in the slow-roll parameters.

\begin{thebibliography}{99}
%
\bibitem{infla}
A.D.~Linde, {\em Particle physics and inflationary cosmology}
(Harwood, Chur, Switzerland, 1990).
A.R.~Liddle and D.H.~Lyth, {\em Cosmological inflation and
large-scale structure} (Cambridge University Press, Cambridge,
England, 2000).
%
\bibitem{wmap}
http://wmap.gsfc.nasa.gov
%
\bibitem{planck}
http://www.rssd.esa.int/index.php?project=Planck
%
\bibitem{SL}
E.D.~Stewart and D.H.~Lyth, Phys. Lett. B 302, 171 (1993).
%
\bibitem{wang_mukh}
L.~Wang, V.F.~Mukhanov and P.J.~Steinhardt,
Phys. Lett. B 414, 18 (1997).
%
\bibitem{SGong}
E.D.~Stewart and J.O.~Gong, Phys. Lett. B 510, 1 (2001).
%
\bibitem{LLMS}
S.M.~Leach, A.R.~Liddle, J.~Martin and D.J.~Schwarz,
Phys. Rev. D 66, 023515 (2002).
%
%
\bibitem{H_MP}
S.~Habib, K.~Heitmann, G.~Jungman and C.~Molina-Par\'{\i}s,
Phys. Rev. Lett. 89, 281301 (2002).
%
\bibitem{HHHJM}
S.~Habib, A.~Heinen, K.~Heitmann, G.~Jungman and
C.~Molina-Par\'{\i}s, Phys. Rev. D 70, 083507 (2004).
%
\bibitem{HHHJ}
S.~Habib, A.~Heinen, K.~Heitmann and G.~Jungman
Phys. Rev. D 71, 043518 (2005).
%
\bibitem{MS_WKB}
J.~Martin and D.J.~Schwarz, Phys. Rev. D 67, 083512 (2003).
%
\bibitem{WKB1}
R.~Casadio, F.~Finelli, M.~Luzzi and G.~Venturi,
Phys. Rev. D 71, 043517 (2005).
%
\bibitem{langer}
R.E.~Langer, Phys. Rev. 51, 669 (1937).
%
\bibitem{MS}
J.~Martin and D.J.~Schwarz, Phys. Rev. D 62, 103520 (2000).
%
\bibitem{WKB_PRL}
R.~Casadio, F.~Finelli, M.~Luzzi and G.~Venturi,
Phys. Lett. B 625, 1 (2005).
%
\bibitem{mukh}
V.F.~Mukhanov, Sov. Phys. JETP Lett. 41, 493 (1985);
Sov. Phys. JETP Lett. 67, 1297 (1988).
%
\bibitem{gris}
L.P.~Grishchuk, Sov. Phys. JETP Lett. 40, 409 (1974).
%
\bibitem{staro}
A.A.~Starobinsky, JETP Lett. 30, 682 (1979).
%
\bibitem{terrero}
D.J.~Schwarz, C.A.~Terrero-Escalante and A.A.~Garc\`{\i}a,
Phys. Lett. B 517, 243 (2001).
%
\bibitem{lewin}
L.~Lewin, \emph{Dilogarithms and associated functions}
(Macdonald, 1958).
%
\bibitem{KT}
A.~Kosowsky and M.S.~Turner,
Phys. Rev. D {\bf 52}, 1739 (1995).
%
\bibitem{LPB}
A.R.~Liddle, P.~Parsons and J.D.~Barrow,
Phys. Rev. D 50, 7222 (1994); see also J.E.~Lidsey, A.R.~Liddle,
E.W.~Kolb, E.J.~Copeland, T.~Barreiro,
and M.~Abney, Rev. Mod. Phys. 69, 373 (1997).
%
\end{thebibliography}
\end{document}